\documentclass[12pt,thmsa,a4paper]{article}
\usepackage{sw20aip}

\input tcilatex
\begin{document}

\title{Algebraic treatment of the confluent Natanzon potentials}
\author{M. T. Chefrour \\
D\'epartement de Physique, Facult\'e des Sciences,\\
Universit\'e Badji Mokhtar, Annaba, Alg\'erie. \and L. Chetouani and L.
Guechi \\
D\'epartement de Physique, Facult\'e des Sciences,\\
Universit\'e Mentouri, Route d'Ain El Bey, \\
Constantine, Alg\'erie.}
\maketitle

\begin{abstract}
Using the $so(2,1)$ Lie algebra and the Baker, Campbell and Hausdorff
formulas, the Green's function for the class of the confluent Natanzon
potentials is constructed straightforwardly. The bound state energy spectrum
is then determined. Eventually, the three-dimensional harmonic potential,
the three-dimensional Coulomb potential and the Morse potential may all be
considered as particular cases.




typescript using Latex (version 2.0).
\end{abstract}

In this paper, we shall want to present an algebraic treatment of the bound
state problem for the class of the confluent Natanzon potentials defined by

\begin{equation}
V(r)=\frac{g_2h^2+g_1h+\eta }R+\frac{\sigma _1h-\sigma _2h^2}{R^2}-\frac 54%
\frac{\Delta h^2}{R^3},  \label{a.1}
\end{equation}
where $\Delta =\sigma _1^2-4\sigma _2c_0$ and $g_1,g_2,\sigma _1,\sigma
_2,c_0$ and $\eta $ are dimensionless parameters. The function $h=h(r)$ is
defined implicitly by the differential equation 
\begin{equation}
\frac{dh}{dr}=\stackrel{.}{h}=\frac{2h}{\sqrt{R(r)}},  \label{a.2}
\end{equation}
where 
\begin{equation}
R(r)=\sigma _2h^2+\sigma _1h+c_0.  \label{a.3}
\end{equation}
These potentials, originally introduced by Natanzon, are interesting in the
sense that the Schr\"odinger equation can be reduced to the confluent
hypergeometric form and they include, as special cases, the radial harmonic
oscillator for $\sigma _2=c_0=0$, the radial Coulomb potential for $\sigma
_1=c_0=0$ and the Morse potential for $\sigma _2=\sigma _1=0.$ These
confluent Natanzon potentials have been recently the object of several
studies. More specifically, Natanzon \cite{GNN} has obtained, in a
Schr\"odinger equation approach, the discrete energy spectrum, the
unnormalized wave functions and has derived the Green's function. Cooper and
al \cite{FCJG} have used the supersymmetric quantum mechanics as an
algebraic method to discuss these potentials. Furthermore, the spectrum of
these potentials has been re-derived by Cordero and Salam\'o \cite{PCSS} ,
still in a Schr\"odinger approach, by making use a particular realization of
the $so(2,1)$ algebraic method which is called spectrum generating algebra
(SGA) \cite{PCGCG} . Finally, this class of the confluent Natanzon
potentials has been solved very recently by Grosche \cite{CG} in the
framework of a Feynman path integral approach.

The Milshtein and Strakhovenko variant (MS) of the $so(2,1)$ algebraic
approach \cite{AIMVMS} provides an elegant and straightforward way of
solving the problem for $V(r)$. As we shall see, this approach consists of
constructing the Green's function by expressing the resolvent operator for
this system in the form of a linear combination of an appropriate
realization of $SO(2,1)$ generators. This form is then written in the
Schwinger's integral representation and the action of the resolvent operator
is obtained by means of two Baker, Campbell and Hausdorff formulas.

Let $G(r,r^{\prime };E)$ be the Green's function associated to the
potentials $U(r)=\frac{\hbar ^2}{2m}V(r)$, which is solution of the
differential equation 
\begin{equation}
\left( H-E\right) G(r,r^{\prime };E)=\frac \hbar i\delta (r-r^{\prime }),
\label{a.4}
\end{equation}
where 
\begin{equation}
H=-\frac{\hbar ^2}{2m}\frac{d^2}{dr^2}+U(r)  \label{a.5}
\end{equation}
is the Hamiltonian of the system and E its energy.

In terms of the function $h(r)$ and by taking into account the relation (\ref
{a.2}), we can write the resolvent operator $H-E$ as follows: 
\begin{eqnarray}
H-E &=&\frac{\hbar ^2}{2m}\left\{ -\frac{d^2}{dr^2}+\frac 14\left( 3\frac{%
\stackrel{..}{h^2}}{\stackrel{.}{h}^2}-2\frac{\stackrel{...}{h}}{\stackrel{.%
}{h}}\right) \right.  \nonumber \\
&&\left. +\frac{\stackrel{.}{h}^2}{4h^2}\left[ (g_2-\sigma _2\epsilon
)h^2+(g_1-\sigma _1\epsilon )h+(\eta -1-c_0\epsilon )\right] \right\} ,
\label{a.6}
\end{eqnarray}
with $E=\frac{\hbar ^2}{2m}\epsilon .$

The point canonical transformation $\xi =h(r)$ leads to 
\begin{equation}
\left( \widetilde{H}-\widetilde{E}\right) \widetilde{G}(\xi ,\xi ^{\prime };%
\widetilde{E})=\frac \hbar i\xi \delta (\xi -\xi ^{\prime }),  \label{a.7}
\end{equation}
where

\begin{equation}
\widetilde{G}(\xi ,\xi ^{\prime };\widetilde{E})=\sqrt{\stackrel{.}{h}(r)%
\stackrel{.}{h}(r^{\prime })}G(r,r^{\prime };E).  \label{a.8}
\end{equation}
Then, the dynamics of the physical system is governed by the new Hamiltonian

\begin{equation}
\widetilde{H}=\frac{\hbar ^2}{2m}\left( -\xi \frac{d^2}{d\xi ^2}+\frac{\eta
-1-c_0\epsilon }{4\xi }+\frac{g_2-\sigma _2\epsilon }4\xi \right) ,
\label{a.9}
\end{equation}
and 
\begin{equation}
\widetilde{E}=-\frac{\hbar ^2}{8m}(g_1-\sigma _1\epsilon )  \label{a.10}
\end{equation}
is the pseudo-energy.

To see that one can introduce the three following generators 
\begin{equation}
T_1(x)=-\frac{\hbar ^2}{2M}\left( \frac{d^2}{dx^2}-\frac{\mu (\mu -1)}{x^2}%
\right) ,T_2(x)=-\frac i2\left( x\frac d{dx}+\frac 12\right) ,T_3(x)=\frac
M{4\hbar ^2}x^2,  \label{a.11}
\end{equation}
satisfying the $so(2,1)$ Lie algebra \cite{BGW} 
\begin{equation}
\left[ T_1,T_2\right] =-iT_1,\quad \quad \left[ T_2,T_3\right] =-iT_3,\quad
\quad \left[ T_1,T_3\right] =-iT_2,  \label{a.12}
\end{equation}
it is convenient to perform the change of variable $\xi =x^2$ and redefine
the Green's function as 
\begin{equation}
\widetilde{G}(\xi ,\xi ^{\prime };\widetilde{E})=\sqrt{xx^{\prime }}%
G(x,x^{\prime };\widetilde{E}).  \label{a.13}
\end{equation}

Under this change the differential equation (\ref{a.7}) becomes

\begin{equation}
\left( T_1(x)+2\hbar ^2\widetilde{\omega }^2T_3(x)-\widetilde{E}\right)
G(x,x^{\prime };\widetilde{E})=\frac \hbar i\delta (x-x^{\prime }),
\label{a.14}
\end{equation}
where 
\begin{equation}
\widetilde{\omega }=\frac \hbar M\left( g_2-\sigma _2\epsilon \right)
^{\frac 12}  \label{a.15}
\end{equation}
with $M=4m.$

Being a linear combination of the generators $T_i$, the operator $\widetilde{%
H}$ shows, as expected, a dynamical symmetry $so(2,1)$. By using the
Schwinger's integral representation \cite{JS} , the solution of the
differential equation (\ref{a.14}) can be written as follows: 
\begin{equation}
G(x,x^{\prime };\widetilde{E})=\int_0^\infty dS\exp \left[ \frac i\hbar
\left( \widetilde{E}+i0\right) S\right] K(x,x^{\prime };S),  \label{a.16}
\end{equation}
where 
\begin{equation}
K(x,x^{\prime };S)=\exp \left\{ -\frac{iS}\hbar \left[ T_1(x)+2\hbar ^2%
\widetilde{\omega }^2T_3(x)\right] \right\} \delta (x-x^{\prime }).
\label{a.17}
\end{equation}
The calculation of this kernel is based upon the use of two Baker,
Campbell-and Hausdorff formulas\cite{BGW}

\begin{equation}
\exp \left\{ -\frac{iS}\hbar \left[ T_1+2\hbar ^2\widetilde{\omega }%
^2T_3\right] \right\} =\exp (-iaT_3)\exp (-ibT_2)\exp (-icT_1),  \label{a.18}
\end{equation}
where 
\begin{equation}
a=2\hbar \widetilde{\omega }\tan (\widetilde{\omega }S),\quad b=2\ln \left[
\cos (\widetilde{\omega }S)\right] ,\quad c=\frac 1{\hbar \widetilde{\omega }%
}\tan (\widetilde{\omega }S),  \label{a.19}
\end{equation}
and 
\begin{equation}
\exp (-i\alpha T_3)\exp (-i\beta T_2)\exp (-i\gamma T_1)=\exp (-icT_1)\exp
(\tau T_3),  \label{a.20}
\end{equation}

with 
\begin{equation}
\alpha =\frac{i\tau }{1-i\tau c/2},\quad \beta =2\ln \left( 1-\frac{i\tau c}%
2\right) ,\quad \gamma =\frac c{1-i\tau c/2}.  \label{a.21}
\end{equation}
It is to be noted that formulas (\ref{a.18}) and (\ref{a.20}) can be easily
checked using the finite representation of $so(2,1)$ Lie algebra where the $%
T_i$ operators are defined in terms of $\sigma _i$ Pauli matrices 
\begin{equation}
T_1=\frac{\sigma _{1-}i\sigma _2}{2\sqrt{2}},\quad T_2=-\frac{i\sigma _3}%
2,\quad T_3=\frac{\sigma _1+i\sigma _2}{2\sqrt{2}}.  \label{a.22}
\end{equation}
Here, we also have to use the Laplace transform of the Dirac distribution 
\begin{equation}
\delta (x-x^{\prime })=\frac M{2\hbar ^2}\frac{x^\mu x^{\prime 1-\mu }}{%
2i\pi }\int_{-i\infty +\delta }^{i\infty +\delta }d\tau \exp \left[ \frac
M{4\hbar ^2}(x^2-x^{\prime 2})\tau \right] ;\delta \prec 0,  \label{a.23}
\end{equation}
in order to obtain a manageable result as follows: 
\begin{equation}
\exp (-i\gamma T_1)x^\mu =\left[ 1-i\gamma T_1+\frac 1{2!}(-i\gamma
T_1)^2+...\right] x^\mu =x^\mu .  \label{a.24}
\end{equation}
Using relations (\ref{a.23}), (\ref{a.18}) and (\ref{a.20}), the kernel (\ref
{a.17}) can now be written 
\begin{mathletters}
\label{a1-2}
\begin{eqnarray}
K(x,x^{\prime };S) &=&\frac M{2\hbar ^2}x^{\prime 1-\mu }\exp (-iaT_3)\exp
(-ibT_2)\frac 1{2i\pi }\int_{-i\infty +\delta }^{i\infty +\delta }d\tau 
\nonumber \\
&&\times \exp \left( -\frac M{4\hbar ^2}x^{\prime 2}\tau \right) \exp
(-icT_1)\exp \left( \frac M{4\hbar ^2}x^2\tau \right) x^\mu  \nonumber \\
\ &=&\frac M{2\hbar ^2}x^{\prime 1-\mu }\exp (-iaT_3)\exp (-ibT_2)\frac
1{2i\pi }\int_{-i\infty +\delta }^{i\infty +\delta }d\tau  \nonumber \\
&&\times \exp \left( -\frac M{4\hbar ^2}x^{\prime 2}\tau \right) \exp
(-icT_1)\exp (\tau T_3)x^\mu  \nonumber \\
\ &=&\frac M{2\hbar ^2}x^{\prime 1-\mu }\exp (-iaT_3)\exp (-ibT_2)\frac
1{2i\pi }\int_{-i\infty +\delta }^{i\infty +\delta }d\tau  \nonumber \\
&&\times \exp \left( -\frac M{4\hbar ^2}x^{\prime 2}\tau \right) \exp
(-i\alpha T_3)\exp (-i\beta T_2)\exp (-i\gamma T_1)x^\mu  \nonumber \\
\ &=&\frac M{2\hbar ^2}x^{\prime 1-\mu }\exp (-iaT_3)\exp (-ibT_2)x^\mu
\frac 1{2i\pi }  \nonumber \\
&&\times \int_{-i\infty +\delta }^{i\infty +\delta }d\tau \frac{\exp \left\{
\frac M{2\hbar ^2}\tau \left( -\frac{x^{\prime 2}}2+\frac{x^2}{2-i\tau c}%
\right) \right\} }{\left( 1-\frac{i\tau c}2\right) ^{\mu +\frac 12}} 
\nonumber \\
\ &=&\frac M{2\hbar ^2}x^{\prime 1-\mu }\exp (-iaT_3)\exp (-ibT_2)x^\mu \exp
\left( \frac{iMx^2}{2\hbar ^2c}\right)  \nonumber \\
&&\times \int_{-i\infty +\delta }^{i\infty +\delta }d\tau \frac{\exp \left( -%
\frac{Mx^{\prime 2}}{4\hbar ^2}\tau \right) \exp \left( \frac{Mx^2}{\hbar
^2c^2}\frac 1{\tau +2i/c}\right) }{\left( 1-\frac{i\tau c}2\right) ^{\mu
+\frac 12}},  \label{a.25}
\end{eqnarray}
where formula\cite{LCLGTFH1,LCLGTFH2} 
\end{mathletters}
\begin{equation}
\exp (-i\beta T_2)f(x)=\exp \left( -\frac \beta 4\right) f\left( e^{-\frac
\beta 2}x\right)  \label{a.25}
\end{equation}
has also been used.

The integral can be calculated thanks to the residue theorem after the $\exp
\left( \frac{Mx^2}{\hbar ^2c^2}\frac 1{\tau +2i/c}\right) $ series has been
effected. Hence we obtain 
\begin{eqnarray}
K(x,x^{\prime };S) &=&\frac M{i\hbar ^2c}\exp (-iaT_3)\exp (-ibT_2)\sqrt{%
xx^{\prime }}\exp \left[ \frac{iM}{2\hbar ^2c}(x^2+x^{\prime 2})\right] 
\nonumber \\
&&\times I_{\mu -\frac 12}\left( \frac{Mxx^{\prime }}{i\hbar ^2c}\right) 
\nonumber \\
&=&\frac M{i\hbar ^2c}\exp \left( -\frac{iaM}{4\hbar ^2}x^2\right) \exp
\left( -\frac b2\right) \sqrt{xx^{\prime }}  \nonumber \\
&&\times \exp \left[ \frac{iM}{2\hbar ^2c}\left( e^{-b}x^2+x^{\prime
2}\right) \right] I_{\mu -\frac 12}\left( \frac{Mxx^{\prime }}{i\hbar
^2ce^{\frac b2}}\right)  \nonumber \\
&=&\frac{M\widetilde{\omega }}{i\hbar \sin (\widetilde{\omega }S)}\sqrt{%
xx^{\prime }}\exp \left[ \frac{iM\widetilde{\omega }}{2\hbar }(x^2+x^{\prime
2})\cot (\widetilde{\omega }S)\right]  \nonumber \\
&&\times I_{\mu -\frac 12}\left( \frac{M\widetilde{\omega }xx^{\prime }}{%
i\hbar \sin (\widetilde{\omega }S)}\right) ,  \label{a.26}
\end{eqnarray}
where 
\begin{equation}
\mu =\frac 12+\sqrt{\eta -c_0\epsilon },  \label{a.27}
\end{equation}
and $I_{\mu -\frac 12}(x)$ is the modified Bessel function.

After inserting (\ref{a.26}) into (\ref{a.16}) and performing the
integration over the time variable $S$ with the help of the following
formula (see Gradshtein and Ryzhik\cite{ISGIMR} , $p.729$ , eq. ($6.669.4$)) 
\begin{eqnarray}
&&\ \ \int_0^\infty dq\frac{e^{-2pq}}{\sinh q}\exp \left[ -\frac
12(x+y)\coth q\right] I_{2\gamma }\left( \frac{\sqrt{xy}}{\sinh q}\right)  
\nonumber \\
\  &=&\frac{\Gamma \left( p+\gamma +\frac 12\right) }{\Gamma (2\gamma +1)%
\sqrt{xy}}M_{-p,\gamma }(x)W_{-p,\gamma }(y),  \label{a.28}
\end{eqnarray}
valid for $Re\left( p+\gamma +\frac 12\right) >0,$ $Re(\gamma )>0$ and $y>x$%
, where $M_{-p,\gamma }(x)$ and $W_{-p,\gamma }(y)$ are the Whittaker
functions, the Green's function (\ref{a.16}) becomes

\begin{equation}
G(x,x^{\prime };\widetilde{E})=\frac{\Gamma \left( p+\frac \mu 2+\frac
14\right) }{2i\widetilde{\omega }\Gamma (\mu +\frac 12)\sqrt{xx^{\prime }}}%
M_{-p,\frac \mu 2-\frac 14}(\frac{M\widetilde{\omega }}\hbar x^{\prime
2})W_{-p,\frac \mu 2-\frac 14}(\frac{M\widetilde{\omega }}\hbar x^2),
\label{a.29}
\end{equation}
where $p=-\frac{\widetilde{E}}{2\hbar \widetilde{\omega }}$ and $x>x^{\prime
}.$

Eventually, by taking into account eqs. (\ref{a.29}), (\ref{a.13}), (\ref
{a.8}), (\ref{a.2}) and the successive transformations $\xi =h(r)=x^2$, the
Green's function associated with the confluent Natanzon potentials may be
written:

\begin{eqnarray}
G(r,r^{\prime };E) &=&\frac{\Gamma \left( -\frac{\widetilde{E}}{2\hbar 
\widetilde{\omega }}+\frac \mu 2+\frac 14\right) }{4i\widetilde{\omega }%
\Gamma (\mu +\frac 12)}\left[ \frac{\sqrt{R(r)R(r^{\prime })}}{%
h(r)h(r^{\prime })}\right] ^{\frac 12}M_{-p,\frac \mu 2-\frac 14}(\frac{M%
\widetilde{\omega }}\hbar h(r^{\prime }))  \nonumber \\
&&\ \ W_{-p,\frac \mu 2-\frac 14}(\frac{M\widetilde{\omega }}\hbar h(r)),
\label{a.30}
\end{eqnarray}
and the bound state energy eigenvalues $E_n=\frac{\hbar ^2}{2m}\epsilon _n$
can be obtained from the poles of $\Gamma \left( -\frac{\widetilde{E}}{%
2\hbar \widetilde{\omega }}+\frac \mu 2+\frac 14\right) $ in (\ref{a.30}), 
\begin{equation}
-\frac{\widetilde{E}}{2\hbar \widetilde{\omega }}+\frac \mu 2+\frac
14=-n,\quad n\in \mathbf{N}_{0.}  \label{a.31}
\end{equation}

Upon substitution of (\ref{a.10}), (\ref{a.15}) and (\ref{a.27}) into (\ref
{a.31}), we see then that 
\begin{equation}
\frac{g_1-\sigma _1\epsilon }{2\sqrt{g_2-\sigma _2\epsilon }}+\sqrt{\eta
-c_0\epsilon }=-(2n+1)  \label{a.32}
\end{equation}
is an equation of fourth degree in $\epsilon _n$ which can be solved after
some simple manipulations \cite{INBKAS} .

Many particular cases of the potentials (\ref{a.1}) arise when some of its
parameters vanish. In the following table, we give the interesting ones,
namely, the radial harmonic oscillator, the radial Kepler-Coulomb potential
and the Morse potential

\[
\begin{tabular}{|l|l|l|l|l|}
\hline
$\sigma _1$ & $\sigma _2$ & $c_0$ & $h(r)$ & $V(r)$ \\ \hline
& $0$ & $0$ & $r^2/\sigma _1$ & $\frac{g_1}{\sigma _1}+\frac{g_2}{\sigma _1^2%
}r^2+\frac{\eta -\frac 14}{r^2}$ \\ \hline
$0$ &  & $0$ & $2r/\sqrt{\sigma _2}$ & $\frac{g_2}{\sigma _2}+\frac{g_1}{2%
\sqrt{\sigma _2}r}+\frac{\eta -1}{4r^2}$ \\ \hline
$0$ & $0$ &  & $e^{-r/\sqrt{c_0}}$ & $\frac{g_2}{c_0}e^{-2r/c_0}+\frac{g_1}{%
c_0}e^{-r/\sqrt{c_0}}+\frac \eta {c_0}$ \\ \hline
\end{tabular}
\]

In conclusion, we have shown that the class of confluent Natanzon potentials
can be described by a particular realization of the $so(2,1)$ which enables
us to work out the Green's function. The realization of the $so(2,1)$
algebra used here differs substantially from that of Cordero and Salam\'o 
\cite{PCSS} . The discrete spectrum is identical with the results obtained
by using either the (SGA) method \cite{PCSS} or the Feynman path integral
approach \cite{CG} .


\begin{thebibliography}{99}
\bibitem{GNN}  G. N. Natanzon, Teor. Mat. Fiz. \textbf{38} (1979) 146.

\bibitem{FCJG}  F. Cooper, J. Ginocchio and A. Kahare, Phys. Rev. \textbf{D
36} (1987) 2458.

\bibitem{PCSS}  P. Cordero and S. Salam\'o, J. Phys. A: Math. Gen. \textbf{24%
} (1991) 5299; Condensed Matter Theories, \textbf{7 }(1992) 49; Found. Phys.%
\textbf{23 }(1993) 675.

\bibitem{PCGCG}  P. Cordero and G. C. Ghirardi, Nuovo Cimento \textbf{A 2}
(1971) 217; Fortschr. Phys. \textbf{20} (1972) 105.

\bibitem{CG}  C. Grosche, J. Phys. A: Math. Gen. \textbf{29} (1996) L183.

\bibitem{AIMVMS}  A. I. Milshtein and V. M. Strakhovenko, Phys. Lett. 
\textbf{A 90} ( 1982 ) 447.

\bibitem{BGW}  B. G. Wybourne,\textit{\ Classical Groups for Physicists
(Wiley, New York, 1974).}

\bibitem{JS}  J. Schwinger, Phys. Rev. \textbf{82} (1951) 664.

\bibitem{LCLGTFH1}  L. Chetouani, L. Guechi and T. F. Hammann, Helv. Phys.
Acta \textbf{65} (1992) 1069.

\bibitem{LCLGTFH2}  L. Chetouani, L. Guechi and T. F. Hammann, J. Math.
Phys. \textbf{33} (1992) 3410.

\bibitem{ISGIMR}  I. S. Gradshtein and I. M. Ryzhik, \textit{Table of
Integrals, Series and Products ( Academic, New York,1965 ).}

\bibitem{INBKAS}  I. N. Bronshtein and K. A. Semendyayev, \textit{A Guide
Book to Mathematics (Harri Deutsch, Frankfurt/Main 1973).}

\bibitem{}  \begin{center}
\end{center}
\end{thebibliography}
\end{document}